\documentstyle[sprocl,twoside]{article}
\newcommand{\be}{\begin{equation}}
\newcommand{\ee}{\end{equation}} \def\la{\mathrel{\mathpalette\fun
<}} 
\def\fun#1#2{\lower3.6pt\vbox{\baselineskip0pt\lineskip.9pt
\ialign{$\mathsurround=0pt#1\hfil
##\hfil$\crcr#2\crcr\sim\crcr}}}

\newcommand{\vew}{\mbox{\boldmath${\rm w}$}}
\newcommand{\vek}{\mbox{\boldmath${\rm k}$}}

\newcommand{\vez}{\mbox{\boldmath${\rm z}$}}

 \newcommand{\lan}{\langle}
 \newcommand{\ran}{\rangle}
\newcommand{\lll}{\langle}
 \newcommand{\rrr}{\rangle}

\begin{document}

\title{THE FEYNMAN-SCHWINGER  (WORLD-LINE)
REPRESENTATION IN PERTURBATIVE QCD}
\author{Yu.A. Simonov$^{1}$ and J.A. Tjon$^{2}$\\
$^1$Jefferson Laboratory, Newport News, USA\\ and\\ 
Institute of Theoretical and Experimental Physics,
Moscow, Russia\\$^2$Institute for Theoretical Physics,University of
Utrecht,\\and\\KVI, University of Groningen,The Netherlands}

\maketitle
\abstracts{
   The proper time path integral representation is derived explicitly
   for an arbitrary $n$-point amplitude in QCD. In the standard
   perturbation theory the formalism allows to sum up the leading
   subseries, e.g. yielding double-logarithm Sudakov asymptotics for
   form factors. Correspondence with the standard perturbation theory is
   established and connection to the Bern-Kosower-Strassler method is
   illustrated.
}

\vspace*{0.8cm}


\newpage 
\section*{Dedication.}
\addcontentsline{toc}{section} {\numberline{}Dedication}

During our studies in field theory we many times returned
to the set of methods and ideas, which was developed by
Misha Marinov. Here also belongs the path integral method in its 
different forms, with spin included in the quantum-mechanical
world-line version, which is described below. The excellent review
written by Misha on the subject which appeared in Physics 
Reports,\cite{0} was a most-read source on the subject at that time. 
  It is a great pleasure and honour for us to dedicate the paper to
     his memory.

\section{Introduction.}

   The present stage of development of field theory in general and of
   QCD in particular requires the exploiting of nonperturbative
   methods in addition to summing up perturbative series. This calls
   for specific methods where dependence on vacuum fields can be made
   simple and explicit. A good example is provided by the so-called
   Fock-Feynman-Schwinger Representation (FSR) based on the
   Fock-Schwinger proper time  and Feynman path integral
   formalism.\cite{1,2} For QED asymptotic estimates the FSR was
   exploited in Ref. 4.
Later on this formalism was rederived 
   in Ref. 5
for scalar quarks in QCD and used in the framework of
   the stochastic background method in.\cite{5}

   More recently some modification of the FSR was suggested in Refs. 7,8.
    The one-loop perturbative amplitudes are especially
   convenient for FSR. These amplitudes were extensively studied in Refs. 9-11.
   Meanwhile the first extension of FSR to nonzero temperature field
   theory was done in.\cite{11,12} This formed
   the basis of a systematic study of the role of
   nonperturbative (NP) configurations in the temperature phase
   transitions.\cite{11,13}

   One of the most important advantages of the FSR is that it allows to
   reduce  physical amplitudes to weighted integrals of 
   averaged Wilson loops. Thus the fields (both perturbative and NP)
   enter only through Wilson loops. For the latter case one can apply
 the cluster expansion method,\cite{14} which allows to sum up a
   series of approximations directly in the exponent.
   As a result one can avoid the summation of Feynman diagrams to get
   the asymptotics of form factors.\cite{15}
   The role of FSR in the treatment of NP effects is more crucial. In
   this case one  can develop a powerful method of background
   perturbation theory\cite{16} treating the NP fields as a 
   background.\cite{17}

   In the present paper the main focus will be on perturbative QCD,
   with the aim of establishing the correspondence between the standard
   perturbative expansion and FSR based expansion, stressing the point that
   FSR allows to make exponentiation in a very simple way. Finally also 
   the relation of FSR to the popular Bern-Kosower-Strassler method 
   is discussed.

   \section{General form of FSR in QCD.}

Let us consider a scalar particle (e.g. Higgs boson) interacting
with the nonabelian vector potential, where the Euclidean Lagrangian
is given by
\be
L_\varphi = \frac12 |D_\mu \varphi|^2 + \frac12 m^2 |\varphi|^2
\equiv \frac12 |(\partial_\mu-ig
A_\mu) \varphi |^2 + \frac12 m^2 |\varphi|^2,
\label{1}
\ee
Using the Fock--Schwinger proper time representation 
the two-point Green's function of $\varphi$ can be written in the
quenched approximation as
\be
G(x,y) = (m^2-D^2_\mu)^{-1}_{xy} = \langle x | P \int^\infty_0 ds
e^{-s(m^2-D^2_\mu)}|y\rangle.
\label{2}
\ee
To obtain the FSR for $G$ a second step is needed.
As in Ref. 2
the matrix element in Eq. (\ref{2})
can be rewritten in the form of a path integral, treating $s$ as the
ordering parameter. Note the difference of the  integral (\ref{2})
from the case of the Abelian QED treated in Refs. 2,4,9:
$A_\mu$ in our case is the matrix operator $A_\mu(x)=
A_\mu^a(x)T^a$. It does not commute for different $x$. Hence
the ordering operator $P$ in Eq. (\ref{2}). The precise meaning of $P$
becomes more clear in the final form of a path integral
\be
G(x,y) =\int^\infty_0\!\! ds (Dz)_{xy}\, e^{-K} P \exp\left(ig \!\!\int^x_y
\!\!A_\mu (z) dz_\mu\right),
\label{3}
\ee
where $K= m^2s+ \frac14 \int^s_0 d\tau (\frac{dz_\mu}{d\tau} )^2$.
In Eq. (\ref{3}) the functional integral can be written as
\be 
(Dz)_{xy} \simeq \lim_{N\to \infty} \prod^N_{n=1}\!\int\!\!
\frac{d^4 z(n)}{(4\pi\varepsilon)^2}\! \int\!\! \frac{d^4p}{(2\pi)^4}
e^{ip \left(\sum^N_{n=1}\! z(n)-(x-y)\right)}
\label{4}
\ee
with $N\varepsilon =s$. 
The last integral in Eq. (\ref{4}) ensures that the path $z_\mu(\tau),
0\leq \tau\leq s$, starts at $z_\mu(0) = y_\mu$ and ends at
$z_\mu(s) =x_\mu$. The form of Eq. (\ref{3}) is the same as in the case
of QED except for the ordering operator $P$ which provides a
precise meaning to the integral of the noncommuting matrices
   $A_{\mu_1}(z_1), A_{\mu_2}(z_2)$ etc. In the case of QCD the forms
   (\ref{3}) and (\ref{4}) were introduced in Refs. 5,6.

The FSR, corresponding to a description in terms of particle
dynamics is equivalent to field theory, when all the
vacuum polarisation contributions are also included,\cite{4}
i.e.
\begin{eqnarray}
&&\sum_{N=0}^{\infty} \frac{1}{N!} \prod_{i=1}^{N} \int\! \frac{ds_i}{s_i}
\!\int\! (Dz_i)_{xx} \exp(-K)\; P\exp\left(ig\! \!\int\limits_y^x
\!\!A_{\mu}(z) dz_{\mu}\right) 
\nonumber\\[1mm]
&&= \int\! D\varphi \exp\left(\!-\!\!\int\! \! d^4 x L_{\varphi}(x)\right ).
\label{bss}
\end{eqnarray}
Both sides are equal to vacuum-vacuum transition amplitude in the presence
of the external nonabelian vector field and hence to each other.
For practical calculations proper regularization of the above equation has
to be done.

The field $A_{\mu}$ in  Eq. (\ref{1}) can be considered
as a classical external field or as a quantum one.
In the latter case the Green's functions $\lll A .. A \rrr$
induce nonlocal current-current interaction terms in the l.h.s. of
Eq. (\ref{bss}).
Such terms can also be generated by the presence of a $\varphi$-field potential,
$V(|\varphi|)$ in the r.h.s. of   Eq. (\ref{bss}). 

The advantage of FSR in this case follows from the very
clear space-time picture of the corresponding dynamics in terms of
particle trajectories.
This is especially important if the currents can be treated
as classical or static (for example, in the heavy quark case).
   The mentioned remark on usefulness of the FSR (\ref{3}) becomes 
clear when one
   considers the physical amplitude, e.g. the Green's function of the
   white state $tr(\varphi^+(x) \varphi(x))$ or its nonlocal version
   $tr[\varphi^+ (x) \Phi(x,y) \varphi(y)]$, where $\Phi(x,y)$ -- to be
   widely used in what follows -- is the parallel transporter along some
   arbitrary contour $C(x,y)$
   \be
   \Phi(x,y) =P\exp\left(ig\!\! \int^x_y \!\!A_\mu(z) dz_\mu\right).
   \label{5}
   \ee
   One has by standard rules
\begin{eqnarray}
 &&G_\varphi (x,y) =\left\langle
    tr\left[\varphi^+(x) \varphi(x)\right]\,
    tr \left[\varphi^+(y)
\varphi(y)\right]\right\rangle_A\nonumber\\[1mm]
 &&
     =\int^\infty_0\!\!ds_1\! \int^\infty_0\!\! ds_2
   (Dz)_{xy}(Dz')_{xy}\, e^{-K-K'} \left\langle W\right\rangle_A + \ldots
   \label{6}
\end{eqnarray}
   where dots stand for the disconnected part,
   $\langle G_\varphi (x,x) G_\varphi (y,y)\rangle_A$. We 
have used the fact that the propagator for the
   charge-conjugated field $\varphi^+$ is proportional to
$\Phi^{\dagger}(x,y) = \Phi(y,x)$.
   Therefore the ordering $P$ must be inverted, $\Phi^{\dagger}(x,y)
   =P \exp (ig \int^y_x A_\mu (z) dz_\mu)$.
    Thus all dependence on $ A_\mu$ in $G_\varphi$ is reduced to the
     Wilson loop average
     \be
     \left\langle W\right\rangle_A=\left\langle tr \,P_C \exp ig \!\!\int_C
 \!\!  A_\mu (z) dz_\mu\right\rangle_A.
     \label{8}
     \ee
      Here $P_C$ is the ordering  around the closed loop
      $C$ passing through the points x and y, the loop
      being made of the paths $z_\mu(\tau)a$,
      $z'_\mu(\tau')$ and to be integrated over.

The FSR can also be used to describe the quark and gluon propagation.
Similar to the QED case, 
      the fermion (quark) Green's function in the presence of an
      Euclidean external gluonic field can be written as
\begin{eqnarray}
     \label{9}
 &&     G_q(x,y)= \langle \psi(x) \bar \psi(y)\rangle_q=\langle x
      |(m_q+\hat D)^{-1}|y\rangle 
\nonumber
\\[1mm]
&&     = \langle x|(m_q-\hat D)(m^2_q-\hat D^2)^{-1}|y\rangle
\nonumber
\\
&&  =  (m_q-\hat D)\!\int^\infty_0\!\! ds (Dz)_{xy} e^{-K} \Phi_\sigma(x,y)\,,
\end{eqnarray}
      where $\Phi_\sigma$ is the same as was introduced in Ref. 2
      \be
      \Phi_\sigma (x,y) = P_A\exp\left(ig \int^x_y A_\mu
      dz_\mu\right)\,P_F 
      \exp \left(g\int^s_0 d\tau \sigma_{\mu\nu} F_{\mu\nu}\right)
      \label{10}
      \ee
      and $\sigma_{\mu\nu}=\frac{1}{4i}
      (\gamma_\mu\gamma_\nu-\gamma_\nu\gamma_\mu)$, while $K$ and
      ($Dz)_{xy}$ are defined in Eqs. (\ref{3}) and (\ref{4}). Note that
      operators $P_A, P_F$ in Eq. (10) preserve the proper ordering of
      matrices $A_\mu$ and $\sigma_{\mu\nu}  F_{\mu\nu}$
      respectively. Explicit examples are considered below.

      Finally we turn to the case of FSR for the valence gluon
      propagating in the background nonabelian field. 
Here we only quote the result for the gluon Green's function in the background
      Feynman gauge.\cite{5,17} We have
      \be
      G_{\mu\nu} (x,y) =
      \langle x|(D^2_\lambda \delta_{\mu\nu}-
      2ig F_{\mu\nu})^{-1}
      |y\rangle
      \label{11}
      \ee
      Proceeding in the same way as for quarks, one obtains the FSR
      for the gluon Green's function
      \be
      G_{\mu\nu} (x,y) =
      \int^{\infty}_0 ds (Dz)_{xy} e^{-K_0}\Phi_{\mu\nu}(x,y),
      \label{12}
      \ee
      where we have defined
\begin{eqnarray}
 K_0&\!\!=&\!\!\frac14 \int^\infty_0 \left
(\frac{dz_\mu}{d\tau}\right)^2
      d\tau, \nonumber\\[1mm]
\Phi_{\mu\nu} (x,y) &\!\!\!\!=&\!\!\!\!\left [P_A\exp
\left(ig\!\!\int^x_y\!\! A_\lambda
      dz_\lambda\right)P_F
\exp \left(2g\!\!\int^s_0\!\! d\tau F_{\sigma\rho}
      (z(\tau))\right)\right]_{\mu\nu}\! .
      \label{13}
\end{eqnarray}
      Now in the same way as is done above for scalars in
Eq.  (\ref{6}), one may consider a Green's function,
corresponding to the physical
      transition amplitude from a white state of $q_1, \bar q_2$ to
      another white state consisting of $q_3,\bar q_4$. It is given by
      \be
      G^\Gamma_{q\bar q} (x,y) =
      \langle G_{q} (x,y)
       \Gamma G_{\bar q} (x,y)\Gamma-
       G_{ q} (x,x)
              \Gamma G_{\bar q} (y,y)\Gamma\rangle_A,
              \label{14}
              \ee
where $\Gamma$ describes the interaction between the $q,\bar{q}$ pair in the
meson. The first term on the r.h.s. of Eq. (\ref{14}) can be
              reduced to the same form as in Eq. (\ref{6}) but with the
              Wilson loop containing ordered insertions of the
              operators $\sigma_{\mu\nu} F_{\mu\nu}$ (cf. Eq.~(\ref{10})).

\section{Perturbation theory in the framework  of FSR.
Identities and partial summation.}

In this section we  discuss in detail how the usual results of
perturbation theory follow from FSR. It  is useful to establish such
a general  connection between the perturbation series (Feynman
diagram technique) and FSR.
At the same time the FSR
presents a unique possibility to sum up Feynman diagrams in a very
simple way, where the final result of the summation is written in an
exponentiated way.\cite{15,17} This method will be discussed  in
the next section.

Consider the FSR for the  quark Green's function. According to
(\ref{9}), the 2-nd order of  perturbative expansion of
Eq. (\ref{10}) can be written as
$$G_q(x,y)= (m_q-\hat D) \!\int^\infty_0\!\!  ds 
\!\int^\infty_0 \!\!  d\tau_1\!
\int^{\infty_1}_0\!\!  d\tau_2\, e^{-K} (Dz)_{xu} d^4u (Dz)_{uv}d^4v(Dz)_{vy}
$$
\be
\times
 \left(ig A_\mu(u)\dot u_\mu+g\sigma_{\mu\nu} F_{\mu\nu} (u)\right)
\left(ig A_\nu(v)\dot v_\nu+ g\sigma_{\lambda\sigma}
F_{\lambda\sigma} (v)\right),
\label{15}
\ee
 where we have used the identities
\be
(Dz)_{xy} = (Dz)_{xu(\tau_1)} d^4u(\tau_1)(Dz)_{u(\tau_1)v(\tau_2)} 
d^4 v(\tau_2) (Dz)_{v(\tau_2)y},
 \label{16}\\[1mm]
\ee
\be 
\int^\infty_0\!\!\!\!\!\!ds\!\!\int^s_0 \!\!\!\!d\tau_1\!\!\int^{\tau_1}_0
\!\!\!\!\!\!d\tau_2
\,f(s,\tau_1,\tau_2)=
\int^\infty_0\!\!\!\!\!\!ds\!\!\int^\infty_0\!\!\!\!
\!\!d\tau_1\!\!\int^\infty_0
\!\!\!\!\!\!d\tau_2
\,f(s+\tau_1+ \tau_2, \tau_1+\tau_2,\tau_2).
 \label{17}
\ee
One can also expand only in the color magnetic moment interaction ($\sigma
F$). This is useful when the spin-dependent interaction can be
treated perturbatively, as it is in most cases for mesons and
baryons (exclusions are Goldstone bosons and nucleons,
where spin interaction is very important and interconnected with
chiral dynamics). In this case one obtains
to the second order in ($\sigma F$)
$$
G_q^{(2)}(x,y) = i(m_q-\hat
D)\!\!\int^\infty_0\!\!ds\!\!\int^\infty_0\!\! d\tau_1\!\!\int^\infty_0
\!\!d\tau_2\, e^{-m^2_q(s+\tau_1+\tau_2)-K_0-K_1-K_2} $$
\be
(Dz)_{xu}\Phi (x,u) g(\sigma F(u)) d^4 u(Dz)_{uv}
\Phi(u,v) g(\sigma F(v)) d^4v(Dz)_{vy}. \label{18} 
\ee
In another way it can be written as
$$ G^{(2)}_q(x,y) =i(m_q-\hat
D) (m^2_q-D^2_\mu)^{-1}_{xu} d^4u~g(\sigma
F(u))(m^2_q-D^2_\mu)^{-1}_{uv} d^4v $$ 
\be 
\times g(\sigma
F(v)) (m^2_q-D^2_\mu)^{-1}_{vy}. \label{19}
\ee
 Here $( m^2_q-D^2_\mu)^{-1}$ is the Green's function of a scalar
 quark in the external gluonic field $A_\mu$.
 This type of expansion is useful also for the study of small-$x$
 behavior of static potential, since
 the correlator $\lan\sigma F(u) \sigma F(v)\ran$ plays an
 important role there.

 However, in establishing the general connection between perturbative
 expansion for Green's functions in FSR and expansions of
 exponential $\Phi_\sigma$ in Eq. (\ref{10}), one encounters a
 technical difficulty since the coupling constant $g$ enters in three
 different ways in FSR:

1. in the factor $(m_q-\hat D)$ in front of the integral in Eq. (\ref{9})

2. in the parallel transporter (the first exponential in Eq. (\ref{10}))

3. in the  exponential of $g(\sigma F)$.

Therefore it is useful to compare the two expansions in the operator form, the
standard one
  $$
  (m+\hat D)^{-1}= (m+\hat \partial - ig \hat A)^{-1}=(m+\hat
  \partial)^{-1}+ (m+\hat \partial)^{-1} ig \hat A (m+\hat
  \partial)^{-1}+
  $$
  \be
  + (m+\hat \partial)^{-1} ig \hat A(m+\hat \partial)^{-1} ig \hat
  A(m+\hat \partial )^{-1}+...
  \label{20}
  \ee
  and the FSR 
  \be
  (m+\hat D)^{-1} = (m-\hat D) (m^2-\partial^2)^{-1}
  \sum^\infty_{n=o}(\delta(m^2-\partial^2)^{-1})^n,
  \label{21}
  \ee
  where we have introduced 
  \be
  \delta=-ig (\hat A\hat \partial+\hat \partial \hat A) - g^2 \hat
  A^2\equiv \hat D^2-\partial^2.
  \label{22}
  \ee
To see how the expansion (\ref{21}) works, using $\hat D=\hat \partial -ig \hat A$
Eq. (\ref{21}) becomes
$$
  (m+\hat D)^{-1}=[(m+\hat \partial)^{-1} +ig \hat A
  (m^2-\partial^2)^{-1}]\sum^\infty_{n=0}
  [\delta(m^2-\partial^2)^{-1}]^n
  $$
Separating out the first term we may rewrite this as
\be
 (m+\hat D)^{-1}=(m+\hat \partial)^{-1} + (m+\hat \partial)^{-1} ig \hat
  A(m-\hat D) (m^2-\partial^2)^{-1}\sum^\infty_{n=0}
     [\delta(m^2-\partial^2)^{-1}]^n
\label{23}
\ee

The last three factors in Eq. (\ref{23}) are the
same as occurring in Eq. (\ref{21}). As a consequence the formal iteration of 
the resulting equation for the Greens' function
reproduces the same series as in Eq. (\ref{20}), showing the equivalence of the
two expansions.

It is important to note that each term in the expansion in powers of
$\delta$, after
 transforming the operator form of Eq. (\ref{21}) into the integral
 form of FSR, becomes an expansion of the  exponential
 $\Phi_\sigma$ in Eq. (\ref{10}) in powers of $g$.
 The second order term
 of this expansion was written down before in Eq. (\ref{15}).

It is our purpose now to establish the connection
between the expansion (\ref{21}), (\ref{23}) and the
expansion of the exponential $\Phi_\sigma$ in 
Eq. (\ref{10}) in the quark propagator (\ref{9}).
One can start with term linear in $\hat A$ and write
(for the Abelian case see Appendix B of Ref. 19)
 \be 
G_q^{(1)} = ig \int G^{(0)}_q(x,z(\tau_1)) d^4
z\frac{\xi_\mu(n)}{\varepsilon}
 \bar A_\mu(\tau_1) G_q^{(0)}(z(\tau_1), y),
\label{24}
\ee
where the notation is clear from the general representation of $G_q$,
given by Eq. (\ref{9})
\be
G_q(x,y)
 =\int^\infty_0 ds e^{-sm^2_q} \prod^N_{n=1}
\frac{d^4\xi(n)}{(4\pi\varepsilon)^2}
\exp\left[-\sum^N_{n=1} \frac{\xi^2(n)}{4\varepsilon} \right ]
\Phi_\sigma (\bar A,\xi) 
\label{25} 
\ee 
with 
$\xi (n) = z(n) - z(n-1),~~ \bar A_\mu (n) =\frac12[A_\mu(z(n))+A_\mu(z(n-1))]$ 
and 
\be 
\Phi_\sigma (\bar A, \xi) =P\exp \{ ig \sum^N_{n=1} \bar
A_\mu(n) \xi_\mu(n) + g\sum^n_{n=1} \sigma_{\mu\nu}
F_{\mu\nu}(z(n))\varepsilon \}.
\label{26} 
\ee 
Representing $\xi(n)$ in Eq. (\ref{24}) as $\frac12(\xi_\mu(L)+\xi_\mu(R))$,
where
 $\xi_\mu(L)$ refers to the integral over $\xi_\mu$ in
 $G_q^{(0)}$
to the left of $\xi_\mu$ in Eq. (\ref{24}) and $\xi_\mu(R)$ to the
integral in $G_q^{(0)}$ standing to the right of $\xi_\mu$, we
obtain 
\be 
\int\xi_\mu(n)\frac{d^4\xi(n)}{(4\pi\varepsilon)^2}
e^{ip\xi-\frac{\xi^2}{4\varepsilon}}=-i\frac{\partial}{\partial
p_\mu}  e^{-ip^2\varepsilon}=2 ip_\mu\varepsilon
e^{-p^2\varepsilon}. \label{27}
\ee
Thus Eq. (\ref{24}) in momentum space becomes 
\be
G_q^{(1)}=-gG^{(0)}_q(q)\lan q|p_\mu A_\mu+A_\mu p_\mu|q'\ran
G^{(0)}_q(q') 
\label{28} 
\ee
In a similar way the second order term from the  coinciding
arguments yields
\be 
G^{(2)}_q(coinc) = -g^2\int G^{(0)}_q(x,z) A^2_\mu(z) d^4 z
G_q^{(0)}(z,y). 
\label{29} 
\ee 
Finally, the first order expansion
of the term $\sigma_{\mu\nu} F_{\mu\nu}$ in Eq. (\ref{10}) yields the
remaining missing component of the combination $\delta$, Eq. (\ref{22}), 
which can be  rewritten as 
\be 
\delta=-ig (A_\mu\partial_\mu+\partial_\mu
A_\mu)-g^2 A^2_\mu+g\sigma_{\mu\nu} F_{\mu\nu}. 
\label{30} 
\ee
Hence the second term in the expansion (\ref{21}) 
\be 
(m+\hat D)^{-1}= (m-\hat D) (m^2-\partial^2)^{-1}+(m-\hat
D)(m^2-\partial^2)^{-1}\delta(m^2-\partial^2)^{-1}+... 
\label{31}
\ee 
is exactly reproduced by the expansion of the FSR (\ref{9}),
where in the first exponential $ \Phi_\sigma$ in Eq. (\ref{9}) one
keeps terms of the first and second order, $O(gA_\mu)$ and
$O((gA_\mu)^2)$, while in the second exponential one keeps only
the first order term $O(g\sigma_{\mu\nu}F_{\mu\nu})$. It is easy
to see that this rule can be generalized to higher orders of
the expansion in $\delta$ in Eq. (\ref{21}) as well.

\section{Summing up leading perturbative contributions in FSR.
Sudakov asymptotics.}

Consider now the n-point Green's function with external momenta $p_i$
at the i-th vertex 
 \be
G(p_1,...p_n)=<J_1(p_1)...J_n(p_n)>, J_i(x)=\psi(x)^{\dag}\Gamma_i\psi(x)
\label{32} 
\ee 
Insertion of Eq. (\ref{9}) into Eq. (\ref{32}) for the
one--fermion loop yields 
\begin{eqnarray}
G(p_1,...p_n)&&=<tr\prod^n_{i=1}\Gamma_i(m_i-\hat D_i)\int^\infty_0
ds_i
\nonumber
\\
&&\times
(Dz^{(i)})_{x^{(i)},x^{(i-1)}}
e^{-K_i}\Phi^{(i)}_\sigma
e^{ip^{(i)} x^{(i)}}dx^{(i)}>_A 
\label{33}
\end{eqnarray}
 We shall disregard in what
follows the factors $\Gamma_i(m_i-\hat D_i)$ since we shall be
interested only in the exponentiated contributions. Performing the
$dx^{(i)}$ integrals, one obtains
 $$ G\to \bar
G_n\delta(\sum^n_{i=1}p_i)(2\pi)^4,
$$
 where
 \be
  \bar
G_n=\int\frac{d^4q}{(2\pi)^4}\prod^n_{i=1}
ds_i\prod^N_{k=1}\frac{d\xi^{(i)}(k)}{(4\pi\varepsilon)^2}
e^{iq^{(i)}\sum_k\xi^{(i)}(k)}e^{-K_i}< \Gamma_i(m_i-\hat
D_i)W_\sigma>
\label{34}
\ee
with
$$ <W_\sigma>=<\prod^n_{i=1}\Phi_\sigma^{(i)}>_A,
$$
where $q^{(i)}$ is the momemtum of the fermion loop going from the
i-th to the (i+1)-th vertex.
The integral $d^4q$ denotes the integral over one of $q^{(i)}$, all
others being expressed through it and all $p_i$.

We note that $<W_\sigma>$ is a gauge invariant quantity summing
all the perturbative exchanges  inside the fixed Wilson contour,
defined by the set $\{\xi^{(i)}(k)\}$. In addition to the usual
Wilson (charge) vertices, there are also magnetic moment vertices
$\sigma F$, hence the notation $<W_\sigma>$.

We concentrate now on the contribution of the $A_\mu$ field in
(\ref{34}), yielding the dominant contribution in the asymptotics
(the reader is referred  for the discussion of the $\sigma
F$ term to the Appendix of Ref. 16.

The  crucial step for what follows is the use of the cluster
expansion method,\cite{14} which yields for $<W_\sigma>\to
<W>$ \be <W>\equiv \exp \{\sum^\infty_{r=1}\frac{(ig)^r}{r!}
\sum_{k_i}\xi_{\mu_1}(k_1)\xi_{\mu_2} (k_2) ... \xi_{\mu_r}
(k_r)\ll A_{\mu_1} (z_{k_1})... A_{\mu_r} (z_{k_r})\gg\}
\label{35} \ee

Here double brackets denote cumulants.\cite{14} The lowest order
contribution (in the exponent) can be expressed through the photon (gluon)
propagator. In the Feynman gauge it is (the gauge is irrelevant
since $<W>$ is gauge invariant) 
\be 
<A_{\mu}(z)A_\nu(z')>=
\frac{\delta_{\mu\nu}C_2(f)\hat 1}{4\pi^2 (z-z')^2}. 
\label{36}
\ee
 Here
$C_2(f)$ is the quadratic Casimir operator for the fundamental
representation, $\hat 1$ is the unit color matrix. For QED one
should replace $C_2\hat 1\to 1$.

  Eq. (\ref{35}) represents the perturbative sum in the exponent,
  which by itself is an important advantage, since each term of
  perturbative expansion is already exponentiated. This property of
  exponentiation is well known for the Wilson loop,
   without using the powerful cluster expansion technique.
In particular, the static
 $Q\bar Q$ potential exponentiates and it can be defined through the
 Wilson loop as follows
 \be
 V(R)=-\lim_{T\to \infty} \frac{1}{T} \ln W(R,T).
 \label{37}
 \ee

In our case of possibly light quarks with masses $m_i$ one should
carry out an additional integration over quark trajectories (i.e. over
$d\xi^{(i)}(k)$). In this section we consider the
asymptotics of the amplitude (\ref{34}) and therefore use
the stationary point analysis for trajectories $\{\xi^{(i)}(k)\}$.
Instead in the next section another method will be exploited,
which was used for partial summation of perturbative diagrams 
in Refs. 9-11.
In that section we shall also show how to
modify this method to include also nonperturbative contributions.

In what follows we confine ourselves to the lowest contribution
(\ref{36}) in Eq. (\ref{35}) and show that it yields the double
logarithmic asymptotics.
First of all one can persuade oneself that the approximation
(\ref{36}) yields in Eq. (\ref{34}) all diagrams with exchanges of
photon/gluon lines between fermion lines,  all orderings of lines
included. For QCD this means the following: all orderings, i.e.
all intersection of gluon lines in space-time are included, except
that the color ordering of operators $t^a$ is kept fixed. Since
the commutator of   any two generators $t^a$ is subleading at large
$N_c$, it means that Eq. (\ref{36}) sums up all exchanges including
intersections of gluonic lines in the leading $N_c$ approximation

Our next point is the integration over $d\xi(k)$ in Eq. (\ref{34}) which is
Gaussian in the main term $K_i$, defining the measure of
integration. Therefore we can carry it out by expanding the  exponent in
Eq. (\ref{34}) around the stable fixed point $\bar \xi$, which is obtained by
differentiating the exponent in (\ref{34}) with respect to
$\xi^{(i)}(k)$. One has
$$
\bar \xi^{(i)}(k)=2\varepsilon_i \{iq^{(i)}-
\frac{g^2C_2(f)}{4\pi^2}\sum_{j,k'} \frac{\bar \xi^{(j)}
(k')}{(\bar z^{(i)}(k)-\bar z^j(k'))^2}+
$$
\be +\frac{2g^2C_2(f)}{4\pi^2}\sum_{j,k'} \sum_{m\geq k}\frac{(
\bar \xi^{(i)} (m)\bar \xi^{(j)}(n'))(\bar z^{(i)}(m)-\bar
z^{(j)}(n'))}{(\bar z^{(i)}(m)-\bar z^{(j)}(n'))^4}\}+ 0(g^4)
\label{38}
 \ee
 Here e.g. $\bar
z^{(i)}(k)=\sum^i_{j=1}\sum^k_{\nu=1}\bar \xi^{(j)}(\nu)$, where
we have chosen as the origin the coordinate $x^{(1)}$ of the first
vertex, and all other coordinates are calculated using the
connection $x^{(i)}-x^{(i-1)}=\sum^N_{k=1}\xi^{(i)}(k)$ with the
cyclic condition $x^{(n+1)}=x^{(1)}$.

One can solve the system of equations (\ref{38}) iteratively  expanding
in powers of $g^2$. The first two terms are given in (\ref{38}), where
one should replace $\bar \xi^{(i)}$ inside the curly brackets by
$2i\varepsilon_i q^{(i)}$. If one represents the exponential
appearing in Eq. (\ref{34}) after insertion of Eq. (\ref{35}) 
as $exp (-f(\xi, q))$, then one can write
\begin{eqnarray}
f(\xi,q) &=&
\sum_{i,k}\frac{(\xi^{(i)}(k))^2}{4\varepsilon_i}
-i\sum_{i,k} q^{(i)} \xi^{(i)} (k)
\nonumber\\
&&-\frac{g^2C_2(f)}{8\pi^2}
\sum_{i,j,kk'}\frac{\xi^i(k) \xi^j(k')}{(z^i(k)-z^j(k'))^2}+0(g^4).
\label{39}
\end{eqnarray}
The Gaussian integration in (\ref{34}) finally yields
 \be \bar G_n\sim
\int\frac{d^4q}{(2\pi)^4} \prod^n_{i=1} ds_i e^{-f(\bar
\xi,q)-\frac{1}{2} tr \ln \varphi},
\label{40}
\ee 
where the matrix $\varphi$ is 
\be
\varphi^{ij}_{kn} = \frac{1}{2} \frac{\partial^2}{\partial
\xi^{(i)}(k) \partial\xi^{(j)} (n)}f(\xi, q)\biggl | _{\xi=\bar
\xi}.
\label{41}
 \ee

The most important  for what follows is the term $f(\bar \xi,q)$
which can be written as (at this point we reestablish Minkowskian
metric)
\be f(\bar \xi, q)
=\sum^n_{i=1}s_i(q^{(i)})^2+\frac{g^2C_2 (f)}{8\pi^2}
\sum_{ij}\int^{s_i}_0 \int^{s_j}_0\frac{d\tau_id\tau_j(q^{(i)}
q^{(j)})}{(\tau_iq^{(i)}-\tau_jq^{(j)} -\Delta_{ij})^2},
\label{42}
 \ee
 where
we have defined $\tau_i=k\varepsilon_i$, and \be
\Delta_{ij}=\sum^{j-1}_{k=i}s_kq^{(k)}, i<j.
\label{43}
 \ee

The integral in the last term on the r.h.s. of Eq. (\ref{42}) can be written
as
 $s_is_j(q^{(i)}q^{(j)})I_{ij}(s,q)$, where
 \be
 I_{ij}(s,q)=\int^1_0\int^1_0\frac{d\alpha
 d\beta}{(\alpha s_iq^{(i)}-\beta
 s_jq^{(j)}-\Delta_{ij})^2}.
 \label{44}
 \ee
 The diagonal terms, $I_{ij},$ with $i=j$ do not contribute to the
 asymptotics and contain only  selfenergy divergencies, which are
 of no interest to us in what follows. Therefore we shall consider
only the nondiagonal terms with $i\neq j$.

 Let us first study the term with $i=j-1$ ("the dressed
 vertex contribution") and $\Delta_{i,i+1}=s_iq^{(i)}$.
 Then Eq. (\ref{44}) is reduced to the form which will be studied
 below
 $$I_i\equiv I_{i,i+1}(s,q)=\int^1_0\int^1_0
 \frac{d\alpha
 d\beta}{(\alpha s_iq^{(i)}+\beta
 s_jq^{(j)})^2}
$$
\be
=\int^1_0\int^1_0
 \frac{d\alpha
 d\beta}{(a^2\alpha^2+\beta^2b^2+2\alpha\beta(ab))}
 \label{45}
 \ee
 with $a=s_iq^{(i)}, b=s_jq^{(j)}, j=i+1$.
 As it stands the integral (\ref{45}) diverges at small $\alpha,\beta$ (or
 at small $\tau_i,\tau_j$ in Eq. (\ref{42})). The origin of this divergence
 becomes physically clear, when one expresses the distance $z^{(i)}$
 from the vertex position,
 (we go over to the Minkowskian space--time)
 \be
 z^{(i)}=2q^{(i)}\tau_i,~~~z^{(j)}=2q^{(j)}\tau_j.
 \label{46}
 \ee
 The quasiclassical motion (\ref{46}) cannot be true for small
 $\tau_i$,
 when quantum fluctuations wash out the straight--line trajectories.
 The lower limit $\tau_{min}$ can be obtained from
  the quantum uncertainty principle
 \be
\Delta z\Delta q\sim (z^{(i)}-z^{(j)})(q^{(i)}-q^{(j)})\sim 1
\label{47}
\ee
Furthermore, we shall be interested in the kinematical region, where
 \be
|q^{(i)}q^{(j)}|\gg(q^{(i)})^2, (q^{(j)})^2.
\label{48}
\ee
 The value of $\tau_{min}$ then
is found from (\ref{47}) to be \be \tau_{min}\sim
\frac{1}{2|q^{(i)}q^{(j)}|}.
\label{49}
 \ee
 Using Eq. (\ref{49}) one can easily
calculate the integral (\ref{45}), since the term $2\alpha\beta(ab)$ in
the denominator of the integrand in Eq. (\ref{45}) always dominates. The
result is 
\be
I_i=\frac{1}{2s_is_{i+1}(q^{(i)}q^{(i+1)})}
ln(2(q^{(i)}q^{(i+1)})s_i)ln(2(q^{(i)}q^{(i+1)})
s_{i+1}).
\label{50}
 \ee
 The integration of the general term $I_{ij}$ with
$j\neq i-1,i+1$ can be done using  the expressions for the Spence
functions.
 However in the
general case the lower limit $\tau_{min}$ is inessential and the
double logarithmic situation does not occur unless there is a
large ratio, $|\frac{(q_iq_k)}{(q_lq_m)}|\gg 1$.

We start with the open triangle, 
corresponding to the Sudakov vertex function asymptotics,
i.e. when the fermion loop is not closed. We have 
\be \bar G_3 = (-i\hat q
+m)^{-1} \Gamma(q,q')(-i\hat q'+m)^{-1}.
\label{51}
 \ee
  In this case there is no
integration over $d^4q$ in Eq. (\ref{34}) and only one integral $I_{12}$
is present in Eqs. (\ref{44}) and (\ref{50}) (we disregard as before the
selfinteracting pieces $I_{ii}$, which do not contribute to the
asymptotics).

Inserting Eq. (\ref{50}) into Eq. (\ref{42}) and
integration over $ds_1ds_2$ in Eq. (\ref{40}) yields the leading
contribution, where in Eq. (\ref{50}) the arguments $s_i$ are taken to
be $s_i = \frac{1}{q_i^2}$.
For the case of QED $(C_2\equiv 1)$ we find
\be 
\Gamma(q,q')\sim \exp (-\frac{\alpha}{2\pi}\ln
\frac{2|qq'|}{q^2} \ln \frac{2|qq'|}{(q')^{2}}),
\label{52}
\ee
which coincides with the known Sudakov asymptotics.

We turn now to the case of QCD, where the basic  triangle diagram
is closed due to color gauge invariance and try to find out
whether the kinematical region (\ref{48}) plays an important role in the
integral over $d^4q$ in Eq. (\ref{40}).

In the general case, when all $q_i$ are unconstrained and
expressed through three external momenta $p_1,p_2,p_3$ and one
integration variable, the region (\ref{48}), yielding double logarithmic
asymptotics (DLA) (\ref{50}), is suppressed due to large values of
$f(\bar \xi, q)$ in the exponent. As a result the integral over
$dq$ does not lead to the DLA form for $\bar G_3$.

The situation changes however, if one considers instead of   $\bar
G_3$ the form factor, i.e. when the pole terms are factored out
from the vertices 2 and 3
and the vertex functions appear
there. To simplify matter, one can consider for the form factor the
same representation (\ref{40}). Under the integral one has the
vertex functions $\psi_i(k_i),i=1,3$, where
$$
k_1=
q^{(1)}+q^{(3)}-p^{(1)}\frac{(q^{(1)}+q^{(3)})p^{(1)}}{(p^{(1)})^2},
$$
\be k_3=
q^{(2)}+q^{(3)}-p^{(3)}\frac{(q^{(2)}+q^{(3)})p^{(3)}}{(p^{(3)})^2}.
\label{53}
 \ee
The definition (\ref{53}) yields in the c.m. system of particle 1
or 3 the
 familiar relative momentum of two emitted fermions. The presence of
 $\psi_i$ imposes a restriction on the momenta $q_i$, namely
 \be
 k^2_1, k^2_3 \la \kappa^2
 \label{54}
 \ee
 where $\kappa^2$ is some hadronic scale.

  Let us define in the Breit system the momenta $$p^{(1)},
  p^{(2)}, p^{(3)}= (p_0,-\frac{\vec Q}{2}),(0,\vec Q), (p_0,
  \frac{\vec Q}{2}),$$ where $\vec Q^2\gg \kappa^2$ and
  $p_0^2=M^2+\frac{\vec Q^2}{4}$.
   One can then easily see, that Eq. (\ref{54}) constrains the region of integration
   over $dq\equiv d^4q_1$ to the region $|\vec q |\sim
   \kappa, |q_0-p_0|\sim \kappa$, and the conditions (\ref{48}) are
   satisfied.
   Hence in this case one recovers the Sudakov asymptotics (\ref{52}), where
   \be
   |qq'|\to |q^{(1)}q^{(2)}|\approx \frac{\vec Q^2}{4},
   q^2\sim q^{\prime 2}\sim \kappa^2,~~
   \alpha\to \alpha_sC_2.
   \label{55}
   \ee

\section{Explicit path integration in  FSR. Connection to the
Bern-Kosower-Strassler method.}

 The path integral (\ref{9}) can be performed in a direct
 way. After the expansion of Eq. (\ref{10}) in powers of $A_\mu$ and
 $F_{\mu\nu}$ and exploiting Fourier transform for the latter one
 obtains Gaussian integrals which can be easily done.

This procedure is similar to  the method introduced in Refs. 9-11
for one-loop diagrams and the effective action in
QED and QCD. To this end we extend the method  of Refs. 9-11
to the case of meson and glueball Green's  functions. We will show, that
 the method used in Refs. 9-11
can be simplified and generalized.

 Let us start with three typical expressions in the order of
 increasing complexity. We may consider the one-loop effective action
  \be
  \Gamma\{A\} =\int^\infty_0 \frac{ds}{s} \xi(s)
  e^{-m^2s-\frac14\int^s_0\dot z^2_\mu d\tau} (Dz)_{xx} tr W(A_\mu),
  \label{56}
  \ee
  where $\xi(s)$ is a regularization. Other expressions, which may be studied,
are the heavy-light quark-antiquark Green's function
  \be
  G_{HL} (Q) =\int^\infty_0 ds
  e^{-m^2s-\frac14\sum^N_{n=1}\frac{\xi^2(n)}{\varepsilon}-iQ\sum^N_{n=1}
  \xi(n)} \prod^N_{n=1}
  \frac{d^4\xi(n)}{(4\pi\varepsilon)^2} \lan W(A_\mu)\ran.
  \label{57}
  \ee
or the Green's function for the arbitrary mass quark-antiquark 
system
$$ G(Q)
=\int^\infty_0 ds_1\int^\infty_0 ds_2 \frac{d^4p}{(2\pi)^4}
e^{-ip_\mu\sum^{N_1}_{n_1=1} \xi^{(1)}_\mu(n_1) -i(Q_\mu-p_\mu)
\sum^{N_2}_{n_2=1} \xi^{(2)}_\mu(n_2)}$$
 \be
\times e^{-m^2_1s_1-m^2_2s_2}\prod^{N_1}_{n_1=1}
\frac{d^4\xi^{(1)}(n_1)}{(4\pi\varepsilon_1)^2} \prod^{N_2}_{n_2=1}
\frac{d^4\xi^{(2)}(n_2)}{(4\pi\varepsilon_2 )^2}  \lan tr
W(A_\mu)\ran .
\label{58} 
\ee 
Here we have introduced  $N_1 \varepsilon_1=s_1, N_2 \varepsilon_2=s_2$.

Let us consider the expression (\ref{57}), since Eq. (\ref{56}) can be
managed easily in the same way.
To proceed one expands in (Eq. \ref{57}) $\lan W(A_\mu)\ran$ in powers
of $A_\mu$ and uses the Fourier transform $\tilde A_\mu{(k)}$ for
the latter. To the second order in $g$ one obtains
$$
G_{HL} (Q) -G^{(0)}_{HL} (Q) =-g^2\int^\infty_0
dse^{-m^2s-iQ\sum\xi(n)-\frac14\sum\frac{\xi^2(n)}{\varepsilon}}
$$
$$\times \prod^N_{n=1}\frac{d\xi(n)}{(4\pi\varepsilon)^2}\int^s_0\dot
z_\mu(t_1)dt_1\int^{t_1}_0 \dot z_\nu(t_2) dt_2\lan\tilde
A_\mu(k_1) \tilde A_\nu (k_2)\ran
$$
 \be
\times\frac{ d^4k^{(1)}
d^4k^{(2)}}{(2\pi)^8}
e^{ik^{(1)}_\mu z_\mu(t_1)+ ik^{(2)}_\mu z_\mu(t_2)} .
 \label{59} 
\ee
Defining similarly to Ref. 11
\be \sum^2_{i=1} ik^{(i)} z(t_i) = i \int^s_0 J_\mu (\tau) z_\mu(\tau)
d\tau,~~ J_\mu(\tau) =k_\mu^{(1)} \delta(\tau-t_1) + k_\mu^{(2)}
\delta(\tau-t_2),
 \label{60} \ee
one arrives at the integral
 \be
I(Q,J)\equiv \prod^N_{i=1} \frac{d^4\xi(i)}{(4\pi\varepsilon)^2}
\exp\{-iQ\sum_n\xi(n)-\frac{1}{4\varepsilon}
\sum_n\xi^2(n)+i\varepsilon\sum^N_{k=1} \xi_\mu(k) \sum^N_{n=k}
J_\mu(n)\},
\label{61} 
\ee 
where we have used that $z_\mu(t)=
\sum^n_{k=1} \xi_\mu(k)$ and interchanged the order of summation
in $\int J_\mu(\tau) z_\mu (\tau) d\tau=\sum J_\mu(n) z_\mu (n)
\varepsilon.$

The Gaussian integration in Eq. (\ref{61}) can be performed trivially and one
obtains 
$$ I(Q,J) =\exp\{ -\varepsilon^3\sum^N_{k=1} \sum^N_{n=k}
\sum^N_{n'=k} J_\mu(n) J_\mu(n') - Q^2s+\varepsilon^2
2Q_\mu\sum^N_{k=1}\sum^N_{n=k} J_\mu(n)\}$$ 
\be 
=\exp
\{-\int^s_0d\tau(\int^s_\tau J_\mu(\tau') d\tau')^2- Q^2s+ 2Q_\mu
\int^s_0d\tau\int^s_\tau J_\mu(\tau') d\tau'\}.
 \label{62} 
\ee
Substituting (\ref{60}) into (\ref{62}) one finally gets 
\be
I(Q,J) =\exp\{-Q^2s+2Q_\mu(k_\mu^{(1)} t_1+k_\mu^{(2)} t_2)
-\sum_{i,j=1,2} k_\mu^{(i)} k^{(j)}_\mu g_B(t_i,t_j)\},
\label{63}
\ee 
where
\be 
g_B(t_i, t_j)=Min (t_i, t_j).
\label{64} 
\ee
In the derivation of Eq. (\ref{63}) one could apply the string vertex
technique  of Refs. 9-11.
A corresponding method for the
Dirac propagator was used in Ref. 20.

So far we have neglected the factors $\dot z_\mu(t_1)\dot
z_\nu(t_2)$ in (\ref{59}). To take them into account one could
exploit the trick suggested in Ref. 11.
Instead we shall use
below another approach, based on the following relation ( we
denote $I_{\mu\nu} (Q,J)$ the integral (\ref{61}) with $\dot z_\mu
\dot z_\nu$ taken into account). Noting that  $\dot z_\mu(t) =
\frac{\xi_\mu (k)}{\varepsilon}$  we obtain
\begin{eqnarray}
I_{\mu\nu}(Q,J)
&\!\!=&\!\!\prod^N_{n=1}\frac{d^4\xi(n)}{(4\pi\varepsilon)^2}\,
\frac{\xi_\mu(k_1)}{\varepsilon}\, \frac{\xi_\nu(k_2)}{\varepsilon}
\nonumber\\
&\!\!\times&\!\!\exp\left[\sum^N_{k=1}\left(i\varepsilon\,\xi_\mu(k)
j_\mu(k) -iQ\,\xi(k) -\frac{\xi^2(k)}{4\varepsilon}\right)
\right],
\label{65} 
\end{eqnarray}
where
$j_{\mu}(k) \equiv \sum^N_{n=k}J_\mu(n)$. Hence one can make a
simple connection 
\be 
I_{\mu\nu} (Q,J) =\left
(\frac{1}{i\varepsilon^2}\frac{\partial}{\partial
j_{\mu}(k_1)}\right ) \left (
\frac{1}{i\varepsilon^2}\frac{\partial}{\partial
j_{\nu}(k_2)}\right ) I(Q,J).
\label{66} 
\ee 
Computing
(\ref{66}) one should take into account that $t_2\leq t_1$.
In doing so one obtains 
\be 
\dot z_\mu^{(1)} \to -
2i(Q_\mu-k_\mu^{(1)});~~ \dot z_\mu^{(2)} \to -
2i(Q_\mu-k_\mu^{(1)}-k^{(2)}_\mu). 
\label{67} 
\ee 
The $HL$
Green's function to the second order $O(g^2)$ has the form
$$
G_{HL}(Q) -G_{HL}^{(0)}(Q)=4g^2\int^\infty_0 dse^{-(m^2+Q^2)s}
\frac{d^4k^{(1)} d^4k^{(2)}}{(2\pi)^8} \lan \tilde A_\mu(k^{(1)})
\tilde A_\nu(k^{(2)})\ran
$$
$$
\times \int^s_0dt_1 \int^{t_1}_0 dt_2(Q_\mu-k^{(1)}_\mu)
(Q_\nu-k^{(2)}_\nu-k_\nu^{(1)})e^{2Q_\mu(k^{(1)}_\mu t_1+k^{(2)}_\mu
t_2)}
$$
\be 
\times\exp [-(k^{(1)})^2t_1-(k^{(2)})^2t_2-2k^{(1)}_\mu k^{(2)}_\mu
t_2]. 
\label{68} 
\ee
In some cases another form is convenient with integrals over $dt_1
dt_2$ taken explicitly,
$$
G_{HL}(Q) -G_{HL}^{(0)}(Q)=4g^2\int \frac{d^4k^{(1)}
d^4k^{(2)}}{(2\pi)^8} \frac{\lan \tilde A_\mu(k^{(1)}) \tilde
A_\nu(k^{(2)})\ran}{2Qk^{(2)}-(k^{(2)})^2-2k^{(1)}k^{(2)}}
$$
$$
\times
\left(Q_\mu\!-\!
k_\mu^{(1)}\!\right)\!\left(Q_\nu-k_\nu^{(2)}-k_\nu^{(1)}\!\right)\left
\{
\frac{1}{2Q(k^{(1)}\!+k^{(2)})-\!(k^{(1)}+k^{(2)})^2}\!
\left[-\frac{1}{m^2+Q^2}\right.\right.$$ 
\be\left.\left.+
\frac{1}{m^2+(Q\!-\!k^{(1)}\!-\!k^{(2)})^2}\right]\!-
\frac{1}{2Qk^{(1)}\!-\!(k^{(1)})^2}\left
[-\frac{1}{m^2\!+\!Q^2}+\frac{1}{m^2\!+\!(Q\!-\!k^{(1)})^2}\right]\right\}.
\label{69} 
\ee

Eqs. (\ref{68}) and (\ref{69}) contain the general result for the
contribution of the second order correlator to $G_{HL}(Q)$. One
can further specify the correlator $\lan\tilde
A_\mu(k^{(1)})\tilde A_\nu(k^{(2)})\ran$. For example, for the
perturbative propagator in the Feynman gauge one has 
\be
\lan\tilde A_\mu(k^{(1)})\tilde
A_\nu(k^{(2)})\ran=\frac{4\pi\delta_{\mu\nu}}{(k^{(1)}-k^{(2)})^2},
\label{69'} 
\ee 
while for the nonperturbative correlator in the
Balitsky gauge one obtains, using results from\,\cite{20}
$$
\lan\tilde A_\mu(k)\tilde A_\nu(k')\ran = \frac{\sigma
e^{-(k_4-k'_4)^2T_g^2}}{\sqrt{\pi} T_g}\int^1_0 dt\int^1_0dt'
\alpha_\mu (t) \alpha_\nu(t')
$$
$$\times (\delta_{\mu\nu} \delta_{ik} -\delta_{i\nu}\delta_{k\mu})
\left (i\frac{\partial}{\partial k_i}i\frac{\partial}{\partial
k'_k}\right) \exp \{ -[(\alpha+\tilde t^{\prime
2})\vek^2+(\alpha+\tilde t^2)\vek^{\prime 2}+
$$
\be
+ 2\tilde t  \tilde t'(\vek\cdot \vek')][4\alpha(\alpha+\tilde
t^2+\tilde t^{\prime 2})]^{-1}\}
\cdot [\alpha(\alpha+\tilde
t^2+\tilde t^{\prime 2})]^{-3/2}.
 \label{69.a}
 \ee
  Here we have
introduced above the regularizing factor $\exp
(-\alpha(\vez^2+\vew^2))$ and used the notation
$$ \tilde t=\frac{t}{2T_g}, \tilde t' =\frac{ t'}{2T_g};~~
\alpha_\mu(t):\alpha_4(t)=1,~ \alpha_i(t)=t,i=1,2,3.$$
It is assumed that the limit $\alpha\to 0$ is taken at the end of
the calculations.

 Inserting Eq. (\ref{69.a}) into Eqs. (\ref{68}) and (\ref{69})
one obtains the lowest order nonperturbative contribution to
$G_{HL}$, which can be studied for all values of $Q$. In
particular, one can expand the result in powers of $1/Q^2$ to
compare with the standard OPE result, where the expansion starts with
$\frac {tr(F_{\mu\nu}(0))^2}{Q4}$. The kernel (\ref{69.a}) is
nonlocal and generates a series in powers of ($QT_g)^2$.
See Ref. 22
for further details.

The equations (\ref{59}-\ref{69}) obtained above are valid in the
second order of expansion of $\lan tr W(A_\mu)\ran$ in $gA_\mu$.
The method used above can easily be generalized to higher orders.
Indeed, writing
$$
\lan tr W(A_\mu)\ran = \sum^\infty_{n=0} (ig)^n \left \lan tr \left \{
\int^s_0A_{\mu_1} (k^{(1)}) \dot z_{\mu_1}(t_1) dt_1
\int^{t_1}_0A_{\mu_2} (k^{(2)}) \dot z_{\mu_2}(t_2) dt_2
\right . \right .
$$
\be
 \left . \left . \ldots
\int^{t_{n-1}}_0A_{\mu_n} (k^{(n)}) \dot z_{\mu_n}(t_n) dt_n
 e^{i\sum^n_{m=1}k^{(m)}z(t_m) } \prod^n_{i=1} \frac{ d^4k^{(i)}}{(2\pi)^4}
\right\}\right\ran,
 \label{72}
\ee
one can define similarly to Eq. (\ref{60}) for the $n$-th term in
Eq. (\ref{72}) 
\be 
J^{(n)}_\mu(\tau) = \sum^n_{i=1} k_\mu^{(i)} \delta (\tau-t_i).
\label{73} 
\ee
Doing the integrals over $\prod^N_{k=1} d\xi(k)$ as before (cf. Eq.
(\ref{61})) one obtains the following result, generalizing
Eq. (\ref{62}), 
\be 
I^{(n)}(Q, J) =\exp\left \{ -\int^s_0\!\! d\tau\left
(\int^s_\tau j_\mu(\tau') d\tau'\right)^2\!\!-Q^2s+2Q_\mu\int^s_0
\!\!d\tau \int^s_\tau \!\!J_\mu(\tau') d\tau'\right\}. \label{74} \ee
Inclusion of the terms $\dot z_{\mu_1} (t_1) ... \dot
z_{\mu_n}(t_n)$ can be done in the same way as in Eq. (\ref{66}) and
the generalization of $I^{(n)}(Q, J)$ to
$I^{(n)}_{\mu_1...\mu_n}(Q,J)$ is straightforward. In this way one
obtains the general form of the expansion of Green's functions
(\ref{56}) -(\ref{58}) to all orders in $g$. In
contrast to the results of Refs. 9-11
one keeps not only 
perturbative but also nonperturbative contributions.

Although we have used here the FSR method to study the case
of scalar quarks, it can also also be applied for quarks with spin starting
from Eq. (\ref{18}). We have shown in the present paper how to use the FSR 
to treat perturbative QCD, establishing in particular the correspondence between
the standard and FSR perturbation expansion.

\addcontentsline{toc}{section} {\numberline{}References}


\begin{thebibliography}{99}
\bibitem{0}
M.S. Marinov, Phys. Rept. {\bf 60C} 1 (1980).
\bibitem{1}
R.P. Feynman, Phys. Rev. {\bf 80} 440 (1950); ibid {\bf 84} 108 (1951).
\bibitem{2}
V.A. Fock, Izvestya Akad. Nauk USSR, OMEN, 1937, p.557; J. Schwinger,
      Phys. Rev. {\bf 82} 664 (1951).
     \bibitem{3} G.A. Milkhin and E.S. Fradkin, ZhETF {\bf 45} 1926 (1963); 
E.S. Fradkin, Trudy FIAN, {\bf 29} 7 (1965).
     \bibitem{4} M.B. Halpern, A. Jevicki and P. Senjovic, Phys. Rev.
     {\bf D16} 2474 (1977);
K. Bardakci and S. Samuel, Phys.Rev. {\bf D18} 2849 (1978);
R. Brandt et al, Phys. Rev. {\bf D19} 1153 (1979); 
S. Samuel, Nucl. Phys. {\bf B149} 517 (1979);
     J. Ishida and A. Hosoya, Progr. Theor. Phys. {\bf 62} 544 (1979).
 \bibitem{5} Yu.A. Simonov, Nucl. Phys. {\bf B307} 512 (1988).
     \bibitem{6} A.I. Karanikas and C.N. Ktorides, Phys. Lett.  {\bf
     B275} 403 (1992); Phys. Rev. {\bf D52} 58883 (1995).
     \bibitem{7} A.I. Karanikas, C.N. Ktorides and N.G. Stefanis, Phys.
      Rev. {\bf D52} 5898 (1995).
      \bibitem{8} Z. Bern and D.A. Kosower, Phys. Rev. Lett. {\bf 66}
      1669 (1991); Nucl. Phys. {\bf B379} 451 (1992).
      \bibitem{9}Z. Bern and D.C. Dunbar, Nucl. Phys. {\bf B379} 562 (1992).
      \bibitem{10} M.J. Strassler, Nucl. Phys. {\bf B385} 145 (1992).
      \bibitem{11} Yu.A. Simonov, JETP Lett {\bf 54} 249 (1991);
ibid. {\bf 55} (1992) 627; Phys.  At. Nucl. {\bf 58} 309 (1995).
\bibitem{12} 
Yu.A. Simonov, In "Varenna 1995, Selected
      topics in nonperturbative QCD", p.319.
\bibitem{13}
      H.G. Dosch, H.-J. Pirner and Yu.A. Simonov, Phys. Lett. {\bf B349}
      335 (1995); E.L. Gubankova and Yu.A. Simonov, Phys.
      Lett. {\bf 360} 93 (1995);
      N.O. Agasyan, in preparation.
\bibitem{14} 
N.G. van Kampen, Phys. Rept. {\bf C24} 171 (1976).
\bibitem{15} 
Yu.A. Simonov, Phys. Lett. {\bf 464 } 265 (1999).
\bibitem{16}
 B.S. De Witt, Phys. Rev. {\bf 162} 1195 1239 (1967);\\
 J. Honerkamp, Nucl. Phys. {\bf B48} 269 (1972);\\
 G.'t Hooft, Nucl. Phys. {\bf B62} 444 (1973); Lectures
at Karpacz, in : Acta Univ. Wratislaviensis {\bf 368} 345 (1976);\\
L.F. Abbot, Nucl. Phys. {\bf B185} 189 (1981).

\bibitem{17} 
Yu.A. Simonov, Phys. At Nucl. {\bf 58} 107 (1995); 
JETP Lett.  {\bf 75} 525 (1993);\\
Yu.A. Simonov, in: Lecture Notes in Physics v.479,
p. 139; ed. H. Latal and W. Schwinger, Springer, 1996.
      
\bibitem{18} Yu.A. Simonov and J.A. Tjon, Ann. Phys. {\bf 228} 1 (1993).

\bibitem{19} 
M. Henneaux and C. Teitelboim, Ann. Phys. {\bf 143} 127 (1982);\\
   V.Ya. Fainberg, A.Va. Marshakov, Nucl.Phys. {\bf B306} 659 (1988).
\bibitem{20} 
Yu.A. Simonov, Phys. At. Nucl. {\bf 60} 2069 (1997), hep-ph/9704301;\\
Yu.A. Simonov and J.A. Tjon, Phys. Rev. {\bf D62} 014501 (2000).

\bibitem{21} 
V.I. Shevchenko and Yu.A. Simonov, hep-ph/0109051.
\end{thebibliography}
\end{document}